\documentclass[
    ,final            % use final for the camera ready runs
  ]
  {aipproc}

\layoutstyle{6x9}

\usepackage{natbib,graphics,graphicx,subfigure}

\newcommand{\etal}{{\sl et al.}}

\newcommand{\JHK}{$J\!H\!K~$}
\newcommand{\degs}{$^{\circ}~$}

\newcommand{\ergs}{$\,$erg$\,$s$^{-1}$}
\newcommand{\Chandra}{{\it Chandra~}}
\newcommand{\arcsec}{$''$}

\newcommand{\ks}{\ensuremath{K_{\rm S}}}
\newcommand{\brg}{{\rm Br-}\ensuremath{\gamma}}
\newcommand{\jh}{\ensuremath{J\!-\!H}}
\newcommand{\jk}{\ensuremath{J\!-\!K_{\rm S}}}
\newcommand{\hk}{\ensuremath{H\!-\!K_{\rm S}}}

\begin{document}

\title{Exploring a New Population of Compact Objects: X-ray and IR Observations of the Galactic Centre}


\author{Reba M. Bandyopadhyay}{
  address={Dept. of Astronomy, University of Florida, Gainesville, FL
  32611 USA} }

\author{Andrew J. Gosling}
 {address={Dept. of Astrophysics, University of Oxford, Oxford OX1
  3RH, UK} 
 ,altaddress={Department of Physical Sciences, P.O. Box 3000, 90014 University of Oulu, Finland} }

\author{Stephen E. Eikenberry}
  {address={Dept. of Astronomy, University of Florida, Gainesville, FL
  32611 USA} }
%  ,altaddress={<author1 address>} % additional visiting address

\author{Michael P. Muno}{
  address={California Institute of Technology, Pasadena, CA 91125 USA}
}
\author{Katherine M. Blundell}
 {address={Dept. of Astrophysics, University of Oxford, Oxford OX1
  3RH, UK} }
\author{Philipp Podsiadlowski}
  {address={Dept. of Astrophysics, University of Oxford, Oxford OX1
  3RH, UK} }
\author{Valerie J. Mikles}
  {address={Dept. of Astronomy, University of Florida, Gainesville, FL
  32611 USA} }
\author{Curtis DeWitt}
  {address={Dept. of Astronomy, University of Florida, Gainesville, FL
  32611 USA} }

\begin{abstract}
I describe the IR and X-ray observational campaign we have undertaken
for the purpose of determining the nature of the faint discrete X-ray
source population discovered by Chandra in the Galactic Center (GC).
Data obtained for this project includes a deep Chandra survey of the
Galactic Bulge; deep, high resolution IR imaging from VLT/ISAAC,
CTIO/ISPI, and the UKIDSS Galactic Plane Survey (GPS); and IR
spectroscopy from VLT/ISAAC and IRTF/SpeX.  By cross-correlating the
GC X-ray imaging from Chandra with our IR surveys, we identify
candidate counterparts to the X-ray sources via astrometry.  Using a
detailed IR extinction map, we are deriving magnitudes and colors for
all the candidates.  Having thus established a target list, we will
use the multi-object IR spectrograph FLAMINGOS-2 on Gemini-South to
carry out a spectroscopic survey of the candidate counterparts, to
search for emission line signatures which are a hallmark of accreting
binaries.  By determining the nature of these X-ray sources, this
FLAMINGOS-2 Galactic Center Survey will have a dramatic impact on our
knowledge of the Galactic accreting binary population.
\end{abstract}

\maketitle

%%%%%%%%%%%%%%%%%%%%%%%%%%%%%%%%%%%%%%%%%%%%
%% MAINMATTER
%%%%%%%%%%%%%%%%%%%%%%%%%%%%%%%%%%%%%%%%%%%%

\section{Introduction}

The unprecedented sensitivity and angular resolution of \Chandra has
been utilized by Wang \etal\ (2002; hereafter W02 \cite{wang}) and
Muno \etal\ (2003; hereafter M03 \cite{muno03}) to investigate the
X-ray source population of the Galactic Center (GC).  The W02 ACIS-I
survey of the central 0.8\degs$\times$2\degs of the GC revealed a
population of $\sim$800 previously undiscovered discrete weak sources
with X-ray luminosities of $10^{32}-10^{35}$\ergs.  M03 imaged the
central 40 pc$^{2}$ (at 8 kpc) around Sgr A*, finding an additional
$\sim$2300 discrete point sources down to a limiting flux of $10^{31}$
erg/s.  More recently, a deeper \Chandra survey of the central
$\sim$1\degs around Sgr A* has been obtained; the combination of all
of these surveys has revealed a total of $>$10,000 discrete sources
(Figure 1; Muno \etal~ 2008, hereafter M08 \cite{muno08}).  The harder
($\geq$3 keV) X-ray sources are likely to be at the distance of the
GC, while the softer sources are likely to be foreground X-ray active
stars or CVs within a few kpc of the Sun.  Some individual sources
have been identified as X-ray transients, high-mass stars, LMXBs, and
CVs.  However, the nature of the majority of these newly detected
sources is as yet unknown.

\begin{figure}
  \includegraphics[width=0.99\textwidth]{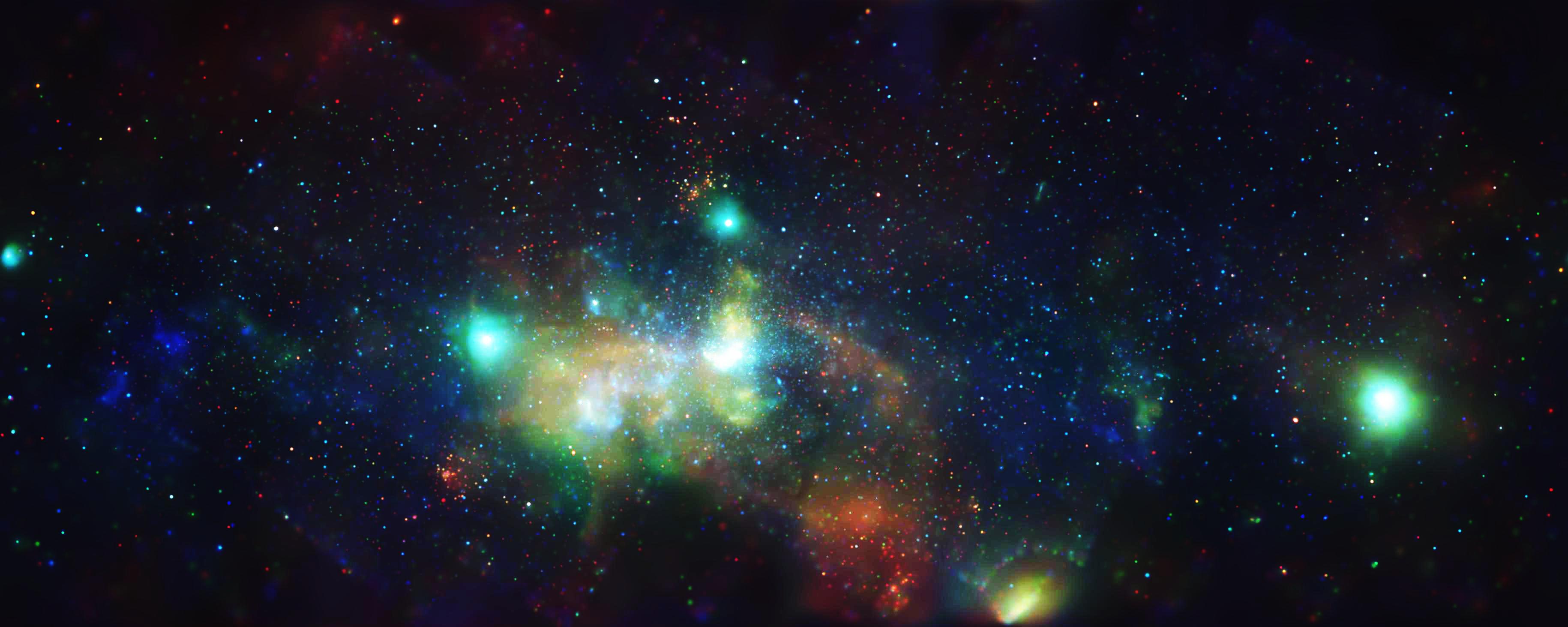} 
  \caption{ACIS-I mosaic of the GC, combining the surveys of W02, M03,
  and M08, as well as additional archival \Chandra data in the region
  \cite{muno08}.  Red is 1-3 keV, green is 3-5 keV, and blue is 5-8
  keV.}
\end{figure}

\section{X-ray/IR Cross-Correlation}

We have undertaken an IR and X-ray observational campaign to determine
the nature of the faint discrete X-ray source population discovered by
\Chandra in the GC.  We have begun by cross-correlating the source
catalog of Muno \etal\ (2006; hereafter M06 \cite{muno06}) with \JHK
images of 26 selected 2.5\arcsec$^{2}$ regions obtained with ISAAC on
the VLT to identify candidate IR counterparts to the X-ray sources
\citep{reba}.

Our original search for the counterparts to the GC X-ray sources used
positions obtained through private communication with the authors of
the W02 survey.  However, M06 re-analyzed the data and produced a
significant refinement of the astrometry of the W02 data; thus the
positions reported for $\sim$50\% of the X-ray sources in the
finalized M06 catalog showed a shift of $\sim 0.5-1$\arcsec\ from the
positions we used for our original cross-correlation analysis.  This
astrometric shift has a substantial impact on the identification of
candidate IR counterparts, since the average stellar separation in our
VLT fields is 1.94\arcsec\ in $K$-band \cite{ajg06}.  In addition, the
revised M06 analysis revealed that a larger number of X-ray sources
were detected in the original \Chandra survey than had been previously
cataloged.  As a result, our VLT imaging survey covers 130 X-ray
sources, compared to the 77 X-ray sources originally reported in
\citet{reba}.

\begin{figure}
    \includegraphics[width=0.99\textwidth]{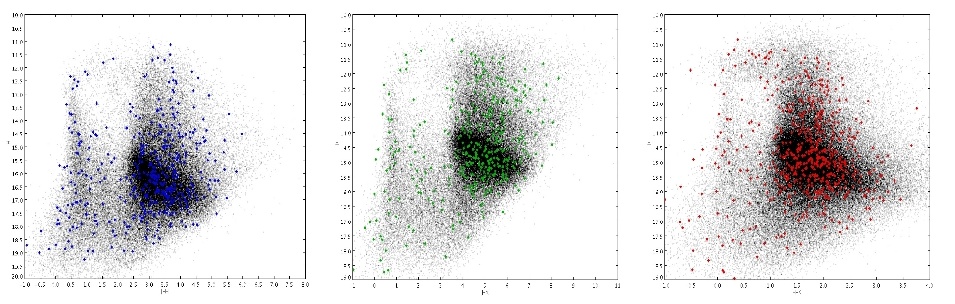}
    \caption{Colour-magnitude diagrams of the candidate counterparts
    to the X-ray sources ({\it coloured points}) and the field stars
    from the VLT Nuclear Bulge observations ({\it black points}). {\it
    Left, blue}: $H\ v\ \jh$; {\it middle, green}: $\ks\ v\
    \jk$; and {\it right, red}: $\ks\ v\ \hk$. The large extension
    towards high values of colour (\textit{right} in each panel)
    indicates that there is a significant amount of near-IR extinction.}
\label{f:colmag}
\end{figure}

Comparing the colours and magnitudes of the candidate counterparts to
the general field populations (Figure 2), we find no clearly
significant difference between the magnitude distribution of candidate
counterparts and those of the field population, aside a slight
over-abundance of more local, bright, blue candidate counterparts (see
especially the left-hand panel of Figure 2).  The colours indicate
that the majority of potential counterparts are located at or beyond
GC distances ($A_{v} \geq 10-30$), with a smaller number consistent
with a foreground stellar population.  $30.8\%$ (40) of the X-ray
sources that lie within the area covered by our VLT observations do
not have detections within the X-ray error circle in any of the three
IR bands.  $58.5\%$ (76) of the X-ray sources had zero candidate
counterpart detections in the $J$-band, compared to $48.5\%$ (63) for
the $H$-band and only $36.2\%$ (47) for the \ks-band.

To examine the significance of these numbers of ``non-matches'', we
performed a Monte Carlo test.  We added a random 0\arcsec --
30\arcsec\ shift to the position of each X-ray source, re-ran the
matching of the X-ray and IR data, and recorded the number of
candidate counterparts identified for the match; we performed this
test 10,000 times, having different random shifts in the X-ray
position for each trial.

The Monte Carlo test revealed that, given randomized X-ray positions,
the expected fraction of X-ray sources that would have no IR
counterparts across all three bands is 41.6$\pm$4.3\% ($1\sigma$
error). For the actual positions, $30.8\%$ of X-ray sources had no
candidate IR counterpart, which is $2.5\sigma$ {\it lower}.  Similarly
for $J$, $H$ and \ks-bands considered individually, the Monte Carlo
simulation revealed that the fraction of X-ray sources for which there
would be no matches if the X-ray sources are randomly scattered across
the fields are 64.8$\pm$4.1\%, 58.3$\pm$4.3\% and 47.5$\pm$4.4\%
respectively. The numbers of ``non-matches'' for the {\it true} X-ray
positions are thus {\it lower} than would be expected from random at
the $1.5\sigma$, $2.3\sigma$ and $2.6\sigma$ levels for $J$, $H$ and
\ks\ respectively.  This result indicates that even with the extremely
high IR source density in the GC, with the corrected and accurate
X-ray astrometry we are now finding significantly {\it more} X-ray/IR
matches than would be expected from chance coincidence.

%%%%%%%%%%%%%%%%%%%%%%%%%%%%%%%%%%%%%%%%%%%%
%% Sample figure:
%%
%% The option [height=...] scales the picture to the given height,
%% without it it would be printed at its nominal size
%%%%%%%%%%%%%%%%%%%%%%%%%%%%%%%%%%%%%%%%%%%%

\section{Extinction}
From our analysis of the VLT images, we find that the IR extinction in
the GC can vary on scales as small as 5\arcsec\ (0.2-0.6 pc at 8 kpc;
\cite{ajg06}).  Some areas show little evidence of this
``granularity'', while others are highly structured.  The relationship
of extinction to wavelength in the IR is a power law with slope
$\alpha$ \citep{mw90}; the ``canonical'' value for $\alpha$ is $\sim$2
\cite{rl85,nish06}.  In contrast, for the GC we find a mean value of
$\alpha = 2.64\pm0.52$; and furthermore, along any given line of sight
to the GC $\alpha$ varies substantially, ranging from $\sim$1.8--3.6
\citep{ajg08}.  We find that the ``universal'' IR extinction law
is {\it not} universal in the GC.  Therefore to obtain
reddening-corrected stellar photometry in the GC, a local value for
the \JHK extinction (on scales $< 20$\arcsec) must be measured and
applied.

\section{Spectroscopy}
The primary accretion signature in the $K$-band which distinguishes a
true X-ray counterpart from a field star is strong Brackett $\gamma$
emission; this technique of identifying XRB counterparts has been
verified with observations of several well-studied GC XRBs
\cite{band99}.  With this technique, using SpeX on IRTF we conclusively
identified a heavily-reddened early-type star as the IR counterpart to
the GC source ``Edd-1'' (CXOGC J174536.1-285638; \cite{edd1,mikles2}).
However, for an additional 16 candidate counterparts for which we
obtained $K$-band spectra with ISAAC, we did not detect \brg\
emission.  A possible explanation is that the accretion signatures
could be too weak to be measureable, for example if the accretion rate
was low at the time of observation, or if the emission was
self-absorbed by the mass donor.  However, a more likely explanation
is that the stars we observed are not the true counterparts to the
X-ray sources.  Our imaging survey had a limiting magnitude of $K$=20,
so our spectroscopic data would only detect XRBs with either
early-type (as in ``Edd-1'') or evolved mass donors.  Therefore it is
likely that the majority of the true IR counterparts belong to a lower
mass population of stars which includes the mass donors of CVs and
LMXBs (as has been suggested by \cite{beltaam04,ebisawa05}, and other
authors); at GC distances, these counterparts are fainter than the
limits of our survey.

\section{Future Work}

Cross-correlation of our IR imaging of the GC with the \Chandra
surveys will produce a large number of IR candidate counterparts to
the X-ray sources.  Due to the extremely high stellar density in the
Nuclear Bulge, many of these astrometric ``matches'' are likely to be
chance superpositions, as indicated by our initial spectroscopic data.
With thousands of candidate counterparts, traditional long-slit
single-target spectroscopy would be a prohibitively inefficient method
by which to identify true counterparts.  Thus we will need to
follow-up with multi-object IR spectroscopy to find the true matches:
this is the work which will be performed with the FLAMINGOS-2 Galactic
Center Survey.

%%%%%%%%%%%%%%%%%%%%%%%%%%%%%%%%%%%%%%%%%%%%%%%%
%% BACKMATTER
%%%%%%%%%%%%%%%%%%%%%%%%%%%%%%%%%%%%%%%%%%%%%%%%

%\begin{theacknowledgments}
%  Infandum, regina, iubes renovare dolorem, Troianas ut opes et
%\end{theacknowledgments}



\bibliographystyle{aipproc}   % if natbib is available

%\bibliography{sample}

%  \typeout{******************************************}
%  \typeout{** Please run "bibtex \jobname" to optain}
%  \typeout{** the bibliography and then re-run LaTeX}
%  \typeout{** twice to fix the references!}
%  \typeout{******************************************}

%\end{document}

%%%%%%%%%%%%%%%%%%%%%%%%%%%%%%%%%%%%%%%%%%%
%% The following lines show an example how to produce a bibliography
%% without the help of the BibTeX program. This could be used instead
%% of the above.
%%%%%%%%%%%%%%%%%%%%%%%%%%%%%%%%%%%%%%%%%%%

\end{document}